\documentclass[twoside, twocolumn]{IEEEtran}

\ifCLASSINFOpdf

\else
\fi

\usepackage{epsfig,tikz,float}
\usepackage{cite}
\usepackage{amsmath}
\usepackage{amssymb}
\usepackage{amsfonts}
\usepackage{algorithmicx}
\usepackage{tikz}
\usetikzlibrary[arrows,graphs,quotes,shapes,chains,trees,positioning,graphs.standard]
\usepackage{algpseudocode}
\usepackage{graphicx}
\usepackage{textcomp}
\usepackage{xcolor}
\usepackage[ruled,vlined,linesnumbered]{algorithm2e}
\usepackage{bm}
\usepackage{bbm}
\usepackage{dsfont}
\usepackage{bbold}
\usepackage{enumitem}
\usepackage{stmaryrd}
\usepackage[utf8]{inputenc} 
\usepackage[T1]{fontenc}
\usepackage{amsthm}
\usepackage[noabbrev]{cleveref}

\usepackage{url}
\usepackage{ifthen}

\newcommand{\E}{\mathsf{E}}

\newcommand{\U}{\mathcal{E}}

\def\BibTeX{{\rm B\kern-.05em{\sc i\kern-.025em b}\kern-.08em
    T\kern-.1667em\lower.7ex\hbox{E}\kern-.125emX}}

\begin{document}
\title{ Feedback Communication Over the BSC with Sparse Feedback times and Causal Encoding\\

\thanks{This work was supported by the National Science Foundation (NSF) under Grant CCF-1955660. Any opinions, findings, and conclusions or recommendations expressed in this material are those of the authors and do not necessarily reflect views of NSF.}

\author{Amaael Antonini\IEEEauthorrefmark{1},
Rita Gimelshein \IEEEauthorrefmark{1},
and Richard D. Wesel\IEEEauthorrefmark{1} 
Email: \{amaael, rgimel,  wesel\}@ucla.edu,  \\

\IEEEauthorblockA{\IEEEauthorrefmark{1}Department of Electrical and Computer Engineering, University of California, Los Angeles, CA 90095, USA} 

}}


\maketitle

\begin{abstract}
Posterior matching uses variable-length encoding of the message controlled by noiseless feedback of the received symbols to achieve high rates for short average blocklengths.  Traditionally, the feedback of a received symbol occurs before the next symbol is transmitted.  The transmitter optimizes the next symbol transmission with full knowledge of every past received symbol. 

To move posterior matching closer to practical communication, this paper seeks to constrain how often feedback can be sent back to the transmitter.  We focus on reducing the frequency of the feedback while still maintaining the high rates that posterior matching achieves with feedback after every symbol.  As it turns out, the frequency of the feedback can be reduced significantly with no noticeable reduction in rate. 

\end{abstract}


\IEEEpeerreviewmaketitle

\section{Introduction}
\label{sec: introduction}

\begin{figure}
\raggedleft
\begin{tikzpicture}
[auto,
		point/.style={circle,inner sep=0pt, minimum size=0.1pt,fill=white},
		skip loop/.style={to path={-- ++(0,#1) -| (\tikztotarget)}},
		decision/.style={circle,draw=black,thick,fill=white,
								text width=0.7em,align=flush center,
								inner sep=0.01pt},
		block/.style={rectangle,draw=black,thick,fill=white,
							text width=3.5em,align=center,rounded corners,
							minimum height=1.5em},
		line/.style={draw,thick,-latex',shorten >=0.1pt},
		cloud/.style={circle,draw=red,thick,ellipse,fill=red!20,
							minimum height=1.5em}
		scale=0.5]
\matrix [column sep=2.5mm, row sep=6.5mm]
{
	&\node 	[block]		(D0)		{Source};
	&\node 	[point]	 	(NULL1)		{};
	&\node 	[block]		(D1)		{Encoder};
	&
	&\node 	[point]	 	(NULL2)		{};
	&
	&\node 	[block] 	(D2) 		{DMC};
        &
	&\node 	[point]	 	(NULL3)		{};
	&
	& \node [block] 	(D3)		{Decoder};
        &
	& \node [point]	 	(NULL4)		{}; 
	&
    \\
	&&
	& \node [point]	 	(NULL5)	{};
	&
	&
	&
	&\node 	[block] 	(D5) 		{Delay};
        &
	& \node [point]	 	(NULL6)	{};
    \\
};
\begin{scope}[every path/.style=line]
\path (D0)--node[above of=NULL1,yshift=-2em]{$\Theta$} (D1);
\path (D0);
\path (D1)--node[above of=D1,yshift=-2em]{$X^{s_l \! + \! D_l}_{s_l \! + \! 1}$}(D2);
\path (D1);
\path (D2)--node[above of=D2,yshift=-2em]{$Y^{s_l \! + \! D_l}_{s_l \! + \! 1}$}  (D3);
\path (D3)--node[above of=D3,yshift=-2em]{$\hat{\Theta}$}  (NULL4);
\path (NULL3)--node [above of=D5,yshift=-2em]{}(NULL6)--node[above of =D5, yshift=-2em] {}(D5);

\path (D5)--node [above of=D1,yshift=-2em]{$Y^{s_l \! + \! D_l}_{s_l \! + \! 1}$}(NULL5)
--node[above of =D1, yshift=-2em] {}(D1);
\end{scope}
\end{tikzpicture}
\vspace{-1.5em}

\caption{System diagram of a BSC with full, noiseless feedback. At sparse times $t=s_1,s_2,\dots, s_\eta$ transmit a block size $D_{l}$}
\vspace{-1.5em}

\label{fig: system model}
\end{figure}
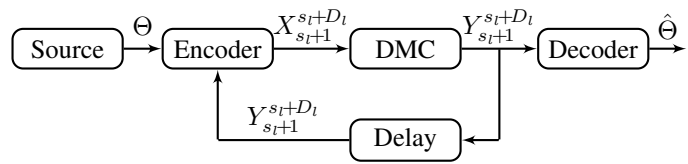

Consider the problem of communicating a $K$-bit message $\Theta$ over a binary symmetric channel (BSC) with a noiseless feedback channel as depicted in Fig. \ref{fig: system model}. At each transmission time $t=1,2, \dots, \tau$ the encoder sends binary symbols $X_t$ through the BSC. The decoder receives symbols $Y_t$ that are noisy versions of $X_t$ where $\Pr(Y_t = 1 \mid X_t = 0) = \Pr(Y_t = 0 \mid X_t = 1) = p$. The receiver may choose to send the symbols $Y_t$ to the transmitter immediately, or allow a few symbols to accumulate, sending all the accumulated symbols in a packet. The receiver needs to produce an estimate $\hat{\Theta}$ of $\Theta$ using the symbols $Y_1,Y_2,\dots,Y_\tau$, and the process ends at the stopping time $\tau$ when the receiver is sufficiently confident of the estimate $\hat{\Theta}$. The goal is to produce the estimate $\hat{\Theta}$ with a low error probability $\Pr(\hat{\Theta}\neq \Theta)$ bounded by a small threshold and with the smallest possible average number of transmissions and average number of feedback transmission instances.

\subsection {Background} 
Shannon \cite{Shannon1956} showed that feedback cannot increase the capacity of discrete memoryless channels (DMC). However, when combined with variable-length coding, Burnashev \cite{Burnashev1976} showed that feedback can help increase the decay rate of the frame error rate (FER) as a function of blocklength. Horstein \cite{Horstein1963} developed one of the earliest schemes for the BSC with noiseless feedback that used sequential transmission and works well in the short blocklength regime. Shayevitz and Feder \cite{Shayevitz2011} later introduced a capacity achieving family of feedback schemes not limited to the BSC, which they called ``posterior matching,'' and showed that it includes Horstein's scheme. 
Li and El-Gamal \cite{Li2015} proposed a fixed length ``posterior matching'' scheme that works well for block-lengths over a few thousand bits.
A notable variable length ``posterior matching'' scheme for a general discrete memoryless channel (DMC) with feedback was proposed by Naghshvar \emph{et. al.}  \cite{Naghshvar2015} and used a sub-martingale analysis to prove that it achieves the channel capacity. 
Other ``posterior matching'' schemes include \cite{Bae2010, Kostina2017, Kim2013, Sabag2018, Truong2014, Anastasopoulos2012}, and more variable length schemes achieving Burnashev's optimal error exponent can be found in \cite{Schalkwijk1971, Schalkwijk1973,Tchamkerten2002,Tchamkerten2006,Naghshvar2012}. 
\thispagestyle{empty}

All previous schemes are sequential schemes where every feedback symbol $Y_t$ is made available to the transmitter before the next symbol $X_{t+1}$ is encoded. In this paper we study the case where the receiver is allowed wait for a few transmissions before sending the accumulated feedback symbols in a single packet. In the meantime, the encoder encodes the next transmissions using only the feedback symbols received in previous feedback packets. Thus, the transmitter can also send those symbols in a single forward transmission packet. We target the short block regime and allow variable feedback transmission intervals.

\subsection{Contributions}
In our precursor journal paper \cite{antonini2023}, we introduced a new analysis for sequential transmission that simplified 
encoding and decoding and improved the rate bound over previous results.
The contributions of the current paper include the following:

\begin{itemize}
\item Show that the same rate bound from \cite{antonini2023} is achievable with packet transmissions instead of sending feedback after every symbol.
\item Introduce new encoding constraints that are less restrictive and better suited for block transmissions.
\item Provide the ``look-ahead'' encoding algorithm that enforces the new encoding constraints for a few transmissions in advance, to allow the transmission of of a packet of symbols, and still guarantees a performance above the lower bound designed for sequential transmission.
\item Provide simulation results that show the achievable feedback sparsity, with an average rate that exceeds the lower bounds developed in \cite{antonini2023}  for sequential transmissions.

\end{itemize}

\subsection{Organization}
\label{sec: organization}

The rest of the paper proceeds as follows. Sec. \ref{sec: sparse system model} describes the sparse feedback times system model, introduces the communication problem and describes the communication scheme by Naghshvar \emph{et. al.}  \cite{Naghshvar2015} on which our methods are based. Sec. \ref{sec: Dense feedback bounds} describes performance bounds for the non-sparse, sequential, feedback model and the encoder that achieves the bound from our previous journal paper \cite{antonini2023}, which we use to benchmark the sparse communication performance.  
Sec. \ref{sec: Sparse Methods} introduces a new encoding constraint that guarantees the bounds in \cite{antonini2023} for sequential feedback but allows sparseness in the feedback times, under certain conditions. Sec. \ref{sec: lookahead algorithm} introduces the ``look-ahead algorithm'' that implements sparse feedback times by encoding 
several symbols in advance, with the guarantee that the constraints in Sec. \ref{sec: Median theorem} will be met for each transmission. Sec. \ref{sec: simulation results} shows the performance of the ``look-ahead algorithm'' in rate, sparsity and complexity from simulations. Sec. \ref{sec: conclusion} concludes the paper.

Throughout the paper we denote random variables (RVs) with upper case letter and instances with lower case letters. We consider discrete times with time indexed by $t=1,2,\dots$. Sequences of random variables  $X_{i}, X_{i+1}, \dots, X_{j}$ will be denoted by $X_{i}^{j}$, possibly dropping the sub index $i$ if $i=1$.

\section{Posterior Matching System Model}
\label{sec: sparse system model}

The system model in Fig. \ref{fig: system model} consists of a source that samples a message $\Theta \in \Omega$ from a distribution $\mathsf{U}(\Omega)$; an encoder that generates the symbols $X_t$ at each time $t$; a discrete memoryless channel that transforms the transmitted symbols $X_t$ into received symbols $Y_t$ according to the channel transition function; a noiseless feedback channel; and a decoder that uses the channel symbols $Y_t$ to produce an estimate $\hat{\Theta}$ of the transmitted message $\Theta$. 

\subsection{Sparse Feedback Times Problem}

The sparse feedback times model allows the receiver to wait a few time indexes between feedback transmission. The received symbols accumulated between feedback transmissions are then sent in a single packet.
The time between feedback transmissions could be variable, just like the block size. Let the feedback transmissions times be at times $t=s_1,s_2,\dots, s_\eta$, with $s_0=0$ and $s_\eta=\tau$. Then, at every time $t=s_{l+1}$ the receiver will send the feedback transmissions corresponding to times  $s_l+1, s_l+2, \dots, s_{l+1}$, in a block of size $D_{l} = s_{l+1}-s_l$, shown by the block $Y^{s_l \! + \! D_l}_{s_l \! + \! 1}$ in Fig. \ref{fig: system model}.

The sparse feedback times communication problem consists of designing a variable length coding scheme to transmit a $K$-bit message using the smallest expected number of channel bits $\tau$ and the smallest number of feedback transmissions $\eta$ that guarantees a frame error rate FER bounded by a small threshold $\epsilon$. We note that the expectations $\E \left[\tau \right]$ and $\E[\eta]$ cannot be minimized at the same time. To see this, note that the minimum of $\E[\eta]$ is zero, which is achieved by any fixed length, forward error correction scheme that guarantees the FER bound. However, as shown by Burnashev \cite{Burnashev1976} feedback and variable rate coding lower the error exponent, which achieves a target FER with a smaller $E[\tau]$. 
To formulate the communication problem  
we need to choose the trade-off between $\E[\tau]$ and $\E[\eta]$. 

There are many ways to formulate the problem to account for the trade-off. One way could be with Lagrange multipliers, where we minimize $\E[\tau]+\lambda \E[\eta]$, for some value of $\lambda$ that could represent the channel access cost, in transmission bits. However, even minimizing $\E[\tau]$ is an integer programming problem whose solution is not yet known to the best of our knowledge.  
Our approach consists of designing a scheme that aims to minimize $\E[\eta]$ while attaining the expected block-length $\E[\tau]$ that satisfies the bound from \cite{antonini2023}. Suppose the bound on $\E[\tau]$ is $\tau_B$, then we can formulate the problem as follows:
\begin{align}
    &\text{minimize }& &\E[\eta]
    \\
    &\text{subject to: }& &\E[\tau] \le \tau_B, \; \Pr(\hat{\theta} \! \neq \! \theta) \le \epsilon
    \label{eq: rate and fer constraints}
    \\    
    &\text{Sparsity Constraint: }& &\mathbf{X}^{s_{l+1}}_{s_l+1} = \mathcal{F}(\theta,\mathbf{y}^{s_l}_1) \, . \label{eq: sparse constraint}
\end{align}
The sparsity constraint restricts the encoder to encode symbols $X_{s_l+2},X_{s_l+3},\dots,X_{s_{l+1}}$ without using the feedback symbols $Y_{s_l+1},Y_{s_l+2},\dots,Y_{s_{l+1}-1}$ not yet re-transmitted by the decoder.
Our approach consists of finding an encoding function that guarantees that constraints \eqref{eq: rate and fer constraints} and \eqref{eq: sparse constraint} are satisfied and seeks to maximize sparsity in the feedback transmission times.

\subsection{Communication Scheme by by Naghshvar \emph{et. al.}}\label{sec: original scheme}
We propose a communication scheme and encoding algorithm that is based on the single phase transmission scheme proposed by Naghshvar \emph{et. al.} \cite{Naghshvar2015}, and combines features of the binary and the non-binary symbols. 
Both encoder and decoder use the channel symbol sequence up to $t$: $\mathbf{Y}^t = Y_1, Y_2,\dots, Y_t$ to compute posterior probabilities $\rho_i(y^t)$ and log likelihood ratio $U_i(t)$ for each possible input message $i$:
\begin{align}
    \rho_i(y^t)&\triangleq P(\theta = i \mid Y^t = y^t), \: \forall i \in \{0,1\}^K \label{eq: postierior defiition}
    \\
    U_i(t) &= U_i(Y^t) \triangleq \log_2 \left( \frac{\rho_i(Y^t)}{1 \! - \! \rho_i(Y^t)}\right)
    \label{eq: U process} \, .
\end{align}
To encode the symbol $X_{t+1}$ the encoder  partitions the message space $\Omega$ into ``bins'', one for each possible input symbol, using a deterministic method known to the decoder. The encoder then transmits the symbol of the bin containing the transmitted message $\theta$. The process terminates once a posterior crosses the threshold $1-\epsilon$ and the message with this posterior is selected as the estimate.
The choice of deterministic partitioning determines the scheme's performance and thus is at the core of the scheme.

In the case of the BSC the encoder by Naghshvar \emph{et. al.} \cite{Naghshvar2015}, needs to construct $2$ sets, $S_0$ and $S_1$, at each time $t$.  To construct the sets,
Naghshvar \emph{et. al.} \cite{Naghshvar2015} proposed a deterministic algorithm, refer to as the ``small enough difference'' encoder (SED) in \cite{Yang2021}, because it satisfies the following constraint:
\begin{equation}
    0\le \sum_{i \in S_0}\rho_i(y^t) - \sum_{i \in S_1}\rho_i(y^t)  < \min_{i \in S_0} \rho_i(y^t) \, .
    \label{eq: sed rule}
\end{equation}
Naghshvar \emph{et. al.} \cite{Naghshvar2015} used the extrinsic Jensen-Shannon  divergence to show that the SED encoder achieves the channel capacity.

\section{Achievable Rate for Sequential Transmission}\label{sec: Dense feedback bounds}
We now describe the best rate lower bound that, to our knowledge, has been developed for sequential transmission over the BSC with noiseless feedback and a simple encoder that achieves it, from our previous work in \cite{antonini2023}. Let $\epsilon$ be the requirement on $\Pr(\hat{\Theta}\neq \Theta)$ and let the block-length be given by the stopping time $\tau$ defined by:
\begin{equation}
    \tau = \min_{t \in \mathrm{N}}\{\exists i \in \Omega: \rho_i(y^t) \ge 1-\epsilon\} \,.
\end{equation}
Let the rate be $K/\E[\tau]$, then a rate lower bound is given by an upper bound on upper bounds on expected block-length $\E[\tau]$. The bound on $\E[\tau]$ from \cite{antonini2023} is given in terms of the channel capacity $C$ and the constants $C_1$ and $C_2$ from \cite{Yang2021}:
\begin{align}
    C &\triangleq 1+p\log_2(p) + (1-p) \log_2(1-p) \label{eq: capacity}
    \\
    C_2 &\triangleq \log_2\left ( \frac{1-p}{p}\right) \label{eq: C2 def}
    \\
    C_1 &\triangleq (1-p) \log_2\left(\frac{1-p}{p}\right) + p \log_2\left(\frac{p}{1-p}\right) \, .
\end{align}
We proposed the ``small enough absolute difference'' (SEAD) encoding rule, a relaxed version of the SED encoder:
\begin{equation}
    -\min_{i \in S_0} \rho_i(y^t) < \sum_{i \in S_0}\rho_i(y^t) - \sum_{i \in S_1}\rho_i(y^t)  \le \min_{i \in S_0} \rho_i(y^t)  \, , \label{eq: SEAD rule}
\end{equation}
and showed that it suffices to guarantee that for all $y^t$ and for some $a > 0$ the following inequalities hold:
\begin{alignat}{4}
    &\E[U_i(t+1)-U_i(t)|\mathcal{F}_t,\theta = j] &&\ge a \label{eq: V ge 0}
    \\
    &U_i(t+1)-U_i(t)   &&\le  C_2  \label{eq: phase 1 max}
    \\
    &E[U_\theta (t+1) - U_\theta (t) | Y^t = y^t] &&\ge C  \, . 
    \label{eq: C step size}
\end{alignat}
If the following singleton constraint is satisfied:
\begin{equation}
    U_i(t) \ge 0 \implies S_0 = \{i\} \; \text{or } S_1 = \{i\} \, , \label{eq: singleton constraint}
\end{equation}
we showed in \cite{antonini2023} that the following inequalities also hold:
\begin{alignat}{4}
    &U_i(t) \ge 0 \implies \E[U_i(t+1)-U_i(t)|\mathcal{F}_t,\theta = j] &&= C_1 \label{eq: phase II V}
    \\
    &U_i(t) \ge 0 \implies \mid U_i(t+1)-U_i(t) \mid  &&= C_2 \, . \label{eq: phase II step size}
\end{alignat}
In \cite{antonini2023} we used a two phase analysis, that divided the transmissions into a communication phase consisting of the times $t$ where $U_\theta(t)\le 0$ with total time $T\triangleq  \sum_{t=1}^\tau \mathbbm{1}_{U_\theta(t) < 0}$, and a confirmation phase with time $\tau-T$. 
We constructed a bound $\tau_{com}$ on $\E[T]$ from inequalities \eqref{eq: V ge 0}
to \eqref{eq: C step size} and a bound $\tau_{conf}$ on $\E[\tau-T]$, with inequalities \eqref{eq: phase II V} and \eqref{eq: phase II step size}, given by:
\ifCLASSOPTIONonecolumn
\begin{align}
    \E[T] &\le \frac{\log_2(M \! - \! 1)+C_2}{C}
    +\left\lceil\frac{\log_2(\frac{1-\epsilon}{\epsilon})}{C_2}\right\rceil\frac{C_2}{C_1} 
    + 2^{-C_2}\left(\frac{ C_2 }{C}-\frac{C_2}{C_1} \right)\frac{1 - \frac{\epsilon}{1-\epsilon}2^{-C_2}}{1 - 2^{-C_2}} \, . 
  \quad   
  \label{eq: stopping time bound}
\end{align}
\else
\begin{flalign}
    &
    \tau_{com}
    \le \frac{\log_2(M \! - \! 1)}{C}+\frac{C_2}{C}
    +2^{-C_2}\frac{C_2}{C}\frac{1 - \frac{\epsilon}{1-\epsilon}2^{-C_2}}{1 - 2^{-C_2}}
    \label{eq: comm phase bound}&
    \\
    &\tau_{conf}
    \le \frac{C_2}{C_1}\left( \! \left\lceil\frac{\log_2(\frac{1\!-\epsilon}{\epsilon})}{C_2}\right\rceil 
    \!- \! 2^{-C_2}\frac{1 \! -\!  \frac{\epsilon}{1\!-\epsilon}2^{-C_2}}{1 - 2^{-C_2}}\!\right) \, . 
  \label{eq: stopping time bound}&
\end{flalign}
\fi
Since $\tau=(\tau-T)+T$, we can construct a bound $\tau_B$ on $\E[\tau]$ using \eqref{eq: comm phase bound} and \eqref{eq: stopping time bound}. However,
bound \eqref{eq: comm phase bound} is loose because of the terms with $\frac{C_2}{C}$ that derive from inequalities \eqref{eq: phase 1 max}
and \eqref{eq: C step size}. 
In \cite{antonini2023} the bound was tighten by constructing a strictly degraded process $U'_i(t)$ that replaced $C_2$ in \eqref{eq: phase 1 max} with $\frac{\log_2(2q)}{q}$, where $q=1-p$. The time $T'$ of the degraded process was lower bounded by that of the original process, that is $T \le T'$. Replacing $C_2$ with $\frac{\log_2(2q)}{q}$ in \eqref{eq: phase 1 max} yields an upper bound $\tau'_{com}$ on $\E[T']$ that applies to both $U_i(t)$ and $U'_i(t)$, given by:
\begin{flalign}
    &
    \tau'_{com}
    \le \! \frac{\log_2(\! M \! \! - \! \! 1\!)}{C} \! + \! \frac{\log_2(2q)}{qC}\! \left(\! \! 1 \! 
    + \! 2^{-C_2}\!\frac{1 \! - \! \frac{\epsilon}{1 \! -\epsilon}2^{-C_2}}{1 - 2^{-C_2}}\! \! \right)
    \label{eq: stopping time optimized bound}&
\end{flalign}

We showed in 
\cite{9174232} and 
\cite{antonini2023} that. when the source samples $\Theta$ uniformly from $\Omega = \{0,1\}^K$, systematic transmissions guarantee that all the constraints are satisfied. At time $t=K$ the posteriors produced by systematic transmissions form a binomial distribution $\mathsf{B}\{0,1\}^K$, which we used in \cite{antonini2023} compute
a bound $\tau^B_{com}$ on $\E[T']$ when $\Theta \sim \mathsf{B}\{0,1\}^K$. Let $\rho_K^h=p^hq^{K-h}$, then $\tau^B_{com}$ is given by:
\ifCLASSOPTIONonecolumn
\begin{align}
    \E[T'-K] 
    \le 
    K + 
    \sum_{i=0}^K \! \left[\frac{\log_2(\frac{1-p^i q^{K\!-\!i}}{p^i q^{K\!-\!i}})}{C} + \frac{\log_2(2q)}{qC}\right] 
    &\binom{K}{i} p^i q^{K-i}  \mathbbm{ 1}_{(q^{K-i}p^i < 0.5)} 
    \nonumber
    \\
    &+
    \left\lceil\frac{\log_2\left(\frac{1-\epsilon}{\epsilon}\right)}{C_2}\right\rceil\frac{C_2}{C_1} 
   \! + \! \left(\frac{\log_2(2q) }{q C} \! - \! \frac{C_2}{C_1} \right)\frac{1 - \frac{\epsilon }{1 \! - \! \epsilon }2^{-C_2}}{1 \! -  \! 2^{-C_2}} 2^{-C_2} \,.  \label{eq: binomial tau}
\end{align}
\else
\begin{flalign}
    \tau^B_{com}  &\le 
    \sum_{h=0}^K \! \left[\frac{\log_2\left(\frac{1\!-\!\rho_K^h}{\rho_K^h}\right)}{C} \! + \! \frac{\log_2(2q)}{qC}\right] \!\! \binom{K}{i} \rho_K^h  \mathbbm{ 1}_{(\rho_K^h < 0.5)} \nonumber& \\
    & \quad \quad \quad \quad + \! \frac{\log_2(2q)}{qC} 2^{-C_2}\!\frac{1 \! - \! \frac{\epsilon}{1 \! -\epsilon}2^{-C_2}}{1 - 2^{-C_2}}\label{eq: binomial T prime}\,.&
\end{flalign}
\fi
In \cite{antonini2023}, we used $\tau=K+(T-K)+(\tau-T)$ to obtain a tighter bound $\tau_B$ on $\E[\tau]$ when $\Theta$ is sampled from a uniform distribution on $\{0,1\}^K$, the bound is given by:
\begin{equation}
    \E[\tau] \le \tau_B = K + \tau^B_{com} + \tau_{conf} \label{eq: binomial tau}
\end{equation}
Bound \eqref{eq: binomial tau} adds $K$ systematic transmissions to the bound $\tau^B_{com}$ on $T'$ for binomial input and the bound $\tau_{conf}$ on $\E[\tau \! - T]$.


\section{Sparse Feedback Times Scheme}
\label{sec: Sparse Methods}
We now show that it is possible to satisfy the constraints  \cref{eq: V ge 0,eq: phase 1 max,eq: C step size} and \eqref{eq: phase II V}, \eqref{eq: phase II step size} to achieve 
an expected stopping time $\E[\tau]$ upper bounded by $\tau_B$ in
\eqref{eq: binomial tau} with some sparsity in the feedback times, i.e. where the feedback is only updated at times $s_1,s_2,\dots$ and not after every transmission. Thus, the transmitter is restricted to encode symbols $X_{s_l+1}, X_{s_l+2}, \dots, X_{s_{l+1}}$ using only the feedback sequence up to time $s_l$, given by $Y_1^{s_l}$.
We will exploit systematic transmissions to make the first feedback time $s_1$ equal to $K$. 

After the systematic transmissions
we will use the non-binary version of the scheme proposed by Naghshvar \emph{et al.} \cite{Naghshvar2015}, where the number of ``bins'' to partition the message set $\Omega$ is the number of symbols in the channel alphabet. We consider 
the block of $D_l$ bits transmitted from time $t=s_l$ to $t=s_{l+1}$ a single symbol out of an alphabet of $2^{D_l}$ symbols, and partitions $\Omega$ into $2^{D_l}$ ``bins.'' The symbol $X_{s_l}^{s_l+D_l}$ transmitted at time $s_l$ will be the $D_l$-bit label assigned to the bin that contains the transmitted message $\theta$, which could just be the index of the bin. The binary partitions  at each transmission $j$ from time $t=s_l$ to $t=s_l+D_l$ will be given by assigning to $S_0$ the ``bins'' whose label has a $0$ at the $j$-th entry to $S_1$ ``bins'' whose label has $1$ at the $j$-th entry. 
Using this scheme, the problem reduces to finding, at each time $s_l$, the largest block size $D_l$ for which we can guarantee that all constraints are met at every time $t=s_l+1, s_l+2,\dots, s_l+D_l$.

\subsection{The ``Weighted Median Absolute Difference'' Rule}
\label{sec: Median theorem}
We now introduce the ``Weighted Median Absolute Difference'' rule, a partitioning rule that further relaxes the tolerance in the difference of sums \eqref{eq: SEAD rule}, sufficient to guarantee constraints \eqref{eq: V ge 0}
to \eqref{eq: C step size}. 
At each time $t$ let $P_0$, $P_1$ and $\Delta$ be:
\begin{align}
    \Delta &\triangleq
    \sum_{i \in S_0}\rho_i(y^t)
    -\sum_{i \in S_1}\rho_i(y^t)
    \\
    P_0 &\triangleq \Pr(\theta \in S_0 \mid Y^t = y^t)
    =\sum_{i \in S_0}\rho_i(y^t) = \frac{1+\Delta}{2} \label{eq: P0 Delta}
    \\
    P_1 &\triangleq \Pr(\theta \in S_1 \mid Y^t = y^t)
    =\sum_{i \in S_1}\rho_i(y^t) = \frac{1-\Delta}{2} \label{eq: P1 Delta}
\end{align}

\begin{figure}[t]
\centering
\ifCLASSOPTIONonecolumn
\includegraphics[width=0.8\textwidth]{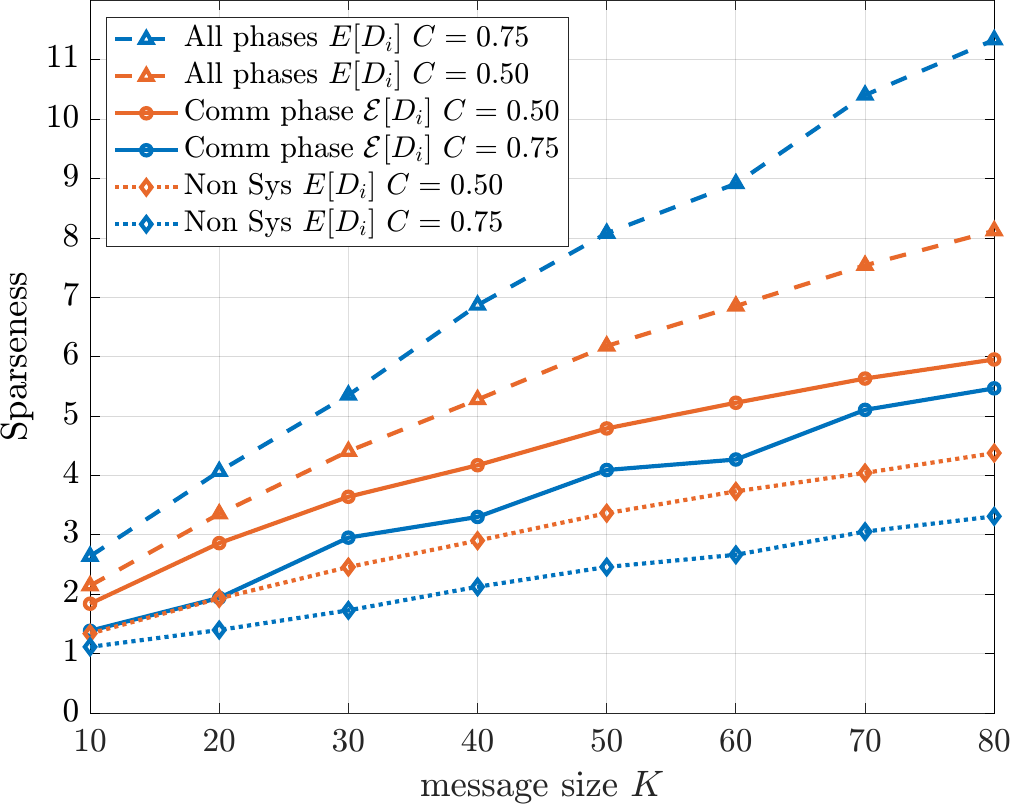}
\else
\includegraphics[width=0.49\textwidth]{Figures/LHsparseVsKC75_50v2.pdf}
\fi
\caption{Feedback sparseness vs. message size $K$ of the ``look-ahead'' algorithm. The curves show average feedback packet size $\E[D_l]$ vs. $K$ for channels with capacity $C=0.50$ and $C=0.75$. The dashed line $--\Delta$ is the overall $\E[D_l]$, the dotted line $\cdot \cdot \Diamond$ excludes the systematic block $D_1=K$ and the solid line $-\circ$ is the performance for only non-systematic transmissions where $\rho_i(y^t) < 0.5 \, \forall i \in \Omega$, which is the target region of the ``look-ahead'' algorithm.}
\label{fig: all sparse}

\vspace{-1.0em}
\end{figure}
Note that $P_0 + P_1=1$, and thus $P_0=\frac{1+\Delta}{2}$ and $P_1=\frac{1-\Delta}{2}$.
Let $\{o_{1},\dots,o_{M}\}$ be an ordering of the vector of posteriors such that $\rho_{o_1}(t) \ge \rho_{o_2}(t) \ge \cdots \ge \rho_{o_M}(t)$, and let $m$ be the index of the ``median'' posterior defined by:
\begin{align}
    \sum_{i=1}^{m-1}\rho_{o_i}(y^t) &< \frac{1}{2}\le \sum_{i=1}^{m}\rho_{o_i}(y^t) \label{eq: sum median}
    \,.
\end{align}
We can show that to satisfy constraints \cref{eq: V ge 0,eq: phase 1 max,eq: C step size,eq: singleton constraint} it suffices to satisfy the following constraint on $\Delta$:
\begin{equation}
    \Delta^2 \le \frac{2}{5}\rho_{o_m}(y^t)  \label{eq: lookahead constraint}
\end{equation}
Rule \eqref{eq: lookahead constraint} offers two significant advantages over SEAD and SED: the first is a larger tolerance on $\Delta$, for most times $s_l$, since $\sqrt{\frac{2}{5}\rho_{o_m}(y^t)}$ is often much larger than $\rho_{o_m}(y^t)$. The second advantage is that the bound on $\Delta$ does not depend on which items are in $S_0$, which allows to allocate items to $S_0$ and $S_1$ to tune $\Delta$ without affecting the tolerance, unlike SED and SEAD in \eqref{eq: sed rule}, \eqref{eq: SEAD rule} where changes in the partitioning cause changes in the tolerance.

To guarantee the bound $\tau_B$ on $\E[\tau]$ in equation \eqref{eq: binomial tau}
we only need to prove that rule \eqref{eq: lookahead constraint} suffices to satisfy constraint \eqref{eq: C step size} and enforce the singleton constraint \eqref{eq: singleton constraint}.
In \cite{antonini2023} we showed that constraint \eqref{eq: phase 1 max} is satisfied by any non-empty $S_0$ and $S_1$, and that $|\Delta| \le 1/3$ suffices to guarantee constraint \eqref{eq: V ge 0}, which can be trivially extended to the any value allowed by rule \eqref{eq: lookahead constraint}.
For the proof see Appendix \ref{sec: delta square proof}


\begin{figure}[t]
\centering
\ifCLASSOPTIONonecolumn
\includegraphics[width=0.8\textwidth]{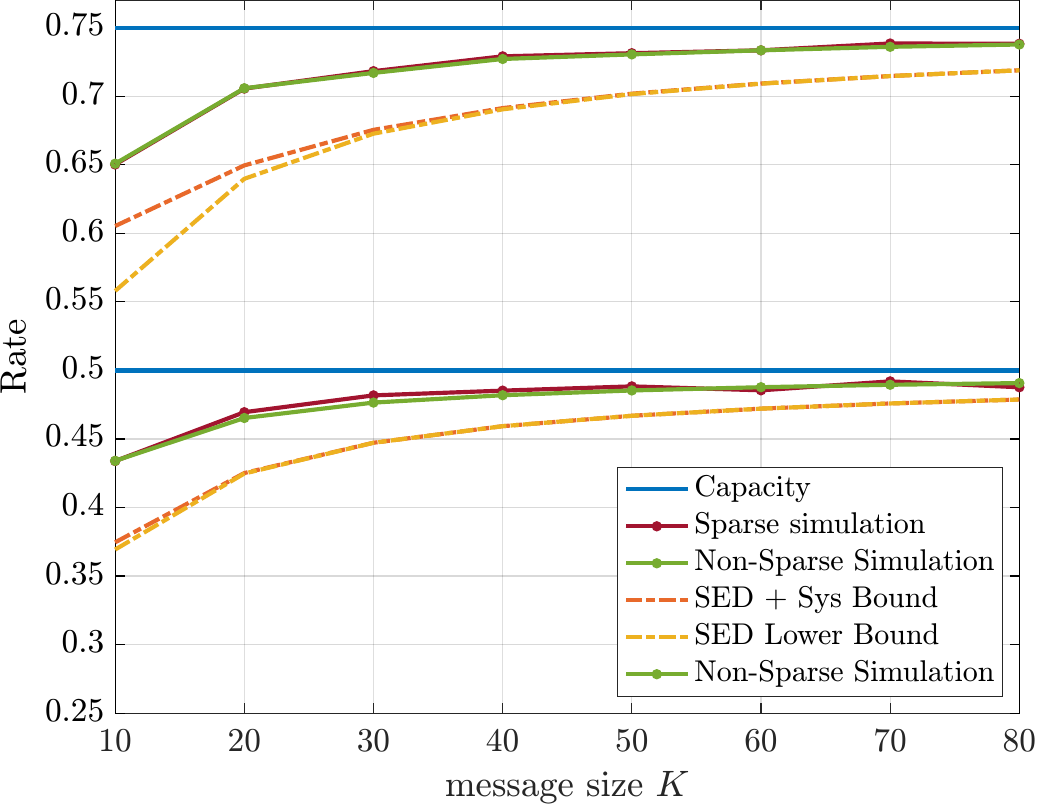}
\else
\includegraphics[width=0.5\textwidth]{Figures/LHrateVsKC75_50v2.pdf}
\fi
\caption{Rate vs message size $K$ for the look-ahead algorithm for two channels with capacities $C=0.50$ and $C=0.75$ shown with the horizontal solid blue lines. The rate performance of the ``look-ahead algorithm'' is shown with the brown solid lines with dots. The green solid line with dots is for a non-sparse algorithm in \cite{antonini2023}. The orange dashed curve is the rate lower bound $K/\E[\tau]$ for for systematic transmission from \eqref{eq: stopping time bound} and the yellow dash dot line is the lower bound from \eqref{eq: stopping time optimized bound} for uniform input distribution.}
\label{fig: rate vs K}

\vspace{-1.0em}
\end{figure}
\subsection{The ``look-ahead'' Algorithm}
\label{sec: lookahead algorithm}
Now we introduce the ``look-ahead'' algorithm, a method to design, based only on $Y_{1}^{s_l}$,
the partitions for the next few transmissions $s_l\!+\!1$, $s_l\!+\!2, \dots, s_l\!+\!D_l$ for some $D_l$. The ``look-ahead'' algorithm needs to guarantee that constraint \eqref{eq: lookahead constraint} is satisfied at each $t\!=\!s_l\!+\!1$, $s_l\!+\!2, \dots, s_l\!+\!D_l$, for the already received sequence $y^{s_l}$ and for each future possible extension sub-sequence $Y_{s_l+1}^{s_l+j}$
, $j=1,2,\dots,D_l-1$. We now identify the key challenges for the ``look-ahead'' algorithm and the steps that we take to overcome these challenges.

First we note that at any time $t=s_l$, only a few $D_l$ values might be feasible, thus, we need to find one such value before designing the partitions. We know from the non-sparse case that $D_l=1$ is always feasible and know how to construct two partitions, say using Naghshvar's algorithm \cite{Naghshvar2015} or the thresholding algorithm 6 in \cite{antonini2023} Sec. VII. 
Second, the algorithm must always converge to a solution in a finite number of steps, which we desire to be reasonably small. For this reason, we will execute a single attempt for a given $D_l$, and upon failure, reduce $D_l$ by one before trying again. This procedure could fall back to the non-sparse case where $D_l=1$.
Third, if we fix $S_0$ and $S_1$ for next times $t=s_l+1,s_l+2,\dots s_l+D_l-1$, then each future $\rho_{o_m}(y^{t})$ and $\Delta$ is a random function of $Y_{s_l+1},Y_{s_l+2},\dots, Y_{s_l + D_l-1}$, the future received symbols. The ``look-ahead'' algorithm needs to guarantee that the pair $\rho_{o_m}(y^{t})$ and $\Delta$ satisfies constraint \eqref{eq: lookahead constraint} at the current time $s_l$ any any future time up to $s_l+D_l-1$.

To overcome these challenges, the  ``look-ahead'' algorithm proceeds as follows: let the $2^{D_l}$ ``bins'' at time $t=s_l$, be $\U_k, \, k=0,1,\dots,2^{D_l}-1$ and define ``bin'' posteriors $P_{\U_k}$, $\delta_k$, and $\delta_{\max}$ by:
\begin{flalign}
    P_{\U_k} &\! \triangleq \! \underset{i \in \U_k}{\sum}\rho_i(y^{s_l}), \,
    \delta_k \! \triangleq \! P_{\U_k} \! \! - \! 2^{-D_l} \! \! , \,
    \delta_{\max} \! \triangleq \! \underset{k}{\max}
    \{|\delta_k|\} ,&
\end{flalign}
where  $2^{-D_l}$ is the target posterior for each bin.
To overcome the uncertainty on $\rho_{o_m}(y^t)$ and guarantee that constraint \eqref{eq: lookahead constraint} on $\Delta$ is satisfied at future time $t=s_l+1, s_l+2,\dots, s_l+D_l-1$ the algorithm finds a lower bound $\rho^{\min}_{o_m}(y^t)$ on $\rho_{o_m}(y^t)$ that is used to compute an upper bound $\Delta_{\max}$ on $\Delta$ for each future time up to $D_l-1$.
The algorithm then uses $\Delta_{\max}$ to determine $\delta_{\max}$ the largest difference $\delta_{k}$ between the posterior $P_{\U_k}$ and the target $2^{-D_l}$.
Note that at each time $s_l+j, \, j=0,1,\dots,D_l-1$ each set $S_x, \, x\in \{0,1\}$ collects the half of ``bins''  whose label has $x$ at entry $j$, Then, $\Delta$ at time $s_l+j$ is given by:
\begin{align}
    |\Delta| = |\sum_{\U_k\in S_0}\delta_k-\sum_{\U_k\in S_1}\delta_k| 
    \le 2^{D_l} \delta_{\max}
\end{align}
Since  $\rho_{o_m}(y^{t+1})$ depend on $\Delta$ at time $t$,
we use an initial $\Delta'_{\max}$ to compute $\rho^{\min}_{o_m}(y^t)$, and then compute bounds $\Delta_{\max}$ on $\Delta$ and $\delta_{\max}$ on each $\delta_k, \, t=s_l, s_l+1,\dots, s_l+D_l-1$ via:
\begin{flalign}
    \Delta_{\max} \! &\triangleq \! \min\{\Delta'_{\max}, \! \sqrt{\frac{2}{5}\rho^{\min}_{o_m}(y^t)}\}, \,
    \delta_{\max} \! \triangleq \! \Delta_{\max}2^{-\! D_l} \! \label{eq: max delta}&
\end{flalign}

We now explain how to compute $\rho^{\min}_{o_m}(y^t)$. Let $x_1^{D_l}(k)$ be the label of bin $\U_k$, and let $Z_k \triangleq \sum_{l=1}^j Y^{s_l\!+\!j}_{s_l+1} \oplus x_1^{D_l}(k)$. At each time 
$t=s_l \! + \! j$ 
the posterior $\rho_i(y^t)$ for $i \in \U_k$ will be:
\begin{flalign}
    \rho_i(y^{s_l \! + \! j}) \! &= \! \frac{\Pr(Y^{s_l\!+\!j}_{s_l+1} \! = \! y^{s_l\!+\!j}_{s_l\!+\!1}  \mid \! Y^{s_l} \! = \! y^{s_l}, \! \theta \! = \! i)\rho_i(y^{s_l})}{
    \sum_{k=0}^{2^{D_l\!-\!1}} \! 
    \Pr(Y^{s_l\!+\!j}_{s_l+1} \! = \! y^{s_l\!+\!j}_{s_l\!+\!1}  \mid \! Y^{s_l} \! = \! y^{s_l}, \! \theta \! \in \! \U_k)P_{\U_k}} \nonumber &
    \\
    &\ge
    \frac{2^jq^{j \! - \! z_k}p^{z_k}\rho_i(y^{s_l})}
    {1 \! + \! \Delta_{\min}}  \ge 
    \frac{2^jq^{j \! - \! z_k}p^{z_k}\rho_i(y^{s_l})}
    {1 \! + \! \Delta'_{\min}}  \label{eq: update bound} 
    \, , &
\end{flalign}
where \eqref{eq: update bound} follows since $\{ Y^{s_l} \! = \! y^{s_l}\}$ determines the partitions $\U_k,\, k=0,1,\dots,2^{D_l}-1$ and $\{\theta \in \U_k\}$ sets $X^{s_l\!+\!j}_{s_l+1} = x_1^j(k)$. 
For the proof of \eqref{eq: update bound} see Appendix \ref{sec: rho tsi min}.
A bound $\rho^{\min}_{o_m}(y^t)$ could just be the smallest item on any collection $\mathcal{C}\subset \Omega$ such that $\sum_{i\in \mathcal{C}} \rho_i(y^t)\ge 1/2$. 
However, we want the largest possible lower bound $\rho^{\min}_{o_m}(y^t)$, thus, we find a collection $\mathcal{C}$ using the items with largest posterior from only the ``bins'' $U_k$ with larger $q^{j \! - \! z_k}p^{z_k}$. We cannot control which bins those will be, but we do know that at time $t=s_l+j$ each value of $z_k$ will be shared by $2^{D_l-j}\binom{j}{z_k}$ ``bins''. Thus we choose $h$, a maximum $z_k$ and find a value $\gamma$ such that: 
\begin{align}
    \gamma 2^{-D_l} \sum^h_{z_k}2^{D_l-j}\binom{j}{z_k}2^jq^{z_k} p^{j-z_k} \ge \frac{1}{2}(1+\Delta'_{\max})\,.
    \label{eq: min gamma}
\end{align}
Now suppose that the posteriors in each bin $\U_k$ are ordered such that $\rho_{o^k_i}(y^{s_l})$ is the $i$-th largest posterior in $\U_k$, and let $\rho_{o^k_\gamma}(y^{s_l})$ be the value of the posterior $\rho_{o^k_n}(y^{s_l})$ such that:
\begin{align}
    \sum_{i=1}^{n-1}\rho_{o^k_i}(y^{s_l}) &< \gamma2^{-D_l} \le \sum_{i=1}^{n}\rho_{o^k_i}(y^{s_l}) \label{eq: bin thresholds}
    \,.
\end{align}
Then, a candidate bound $\rho^{\min}_{o_m}(y^{t})$ on $\rho_{o_m}(y^t)$ at time $t=s_l+j$ is given by the smallest value of $\rho_{o^k_\gamma}(y^{s_l})$, with the worst coefficient $q^{j-h}p^h$, given by:
\begin{align}
    \rho^{\min}_{o_m}(y^{s_l+j}) \triangleq 2^{j-h}p^h \min_{k=0,1,\dots,2^{D_l}\!-\! 1}\{\rho_{o^k_\gamma}(y^{s_l})\} \label{eq: min rho gamma k}
\end{align}

Two steps can help keep the right side of \eqref{eq: min rho gamma k} as large as possible. The first is by making the smallest $\rho_{o^k_\gamma}(y^{s_l})$ the largest possible, and the second is choosing the best possible value of $h$.
We describe first how to make the smallest $\rho_{o^k_\gamma}(y^{s_l})$ as large as possible for a fix $\gamma$.

Let $\rho_{o_\gamma}(y^{s_l})$ be defined in the same manner as $\rho_{o_m}(y^{s_l})$ from \eqref{eq: sum median}, but using $\gamma$ instead of $\frac{1}{2}$ as follows:
\begin{flalign}
    &\rho_{o_\gamma}(y^{s_l}\!) \triangleq \rho_{o_k}(y^{s_l}\!) \; \text{s.t. }\sum_{i=1}^{k-\!1}\!\rho_{o_i}(y^{s_l}\!) \! < \! \gamma \! \le \! \sum_{i=1}^{m}\!\rho_{o_i}(y^{s_l}\!)
    \label{eq: rho gamma}
    \,.&
\end{flalign}
If we distribute the items $i$ with $\rho_{o_i}(y^{s_l})\ge \rho_{o_\gamma}(y^{s_l})$ evenly across all the ``bins,'' then all bins will contain at exactly $\gamma 2^{-D_l}$ and at least one bin will contain an item $i$ with  $\rho_{i}(y^{s_l})= \rho_{o_\gamma}(y^{s_l})$. Since the ``bins'' contain an integer number of items, this might not be possible. However, we can allow a bin to cross $\gamma 2^{-D_l}$ only when the next largest item we allocate does not fit in any other ``bin,'' until each bin crosses $\gamma 2^{-D_l}$. Only then we allocate the smaller items in any order. We still need to make sure that each $\delta_k \le \delta_{\max}$, otherwise the partitioning fails.

Now we describe how to choose $h$ and $\gamma$. Instead of choosing $h$ and computing $\gamma$, we choose $\gamma$ and then find the smallest $h$ that satisfies \eqref{eq: min gamma}. The smallest the value of $\gamma$, the larger the value of $h$. We first show that $\gamma > \frac{1}{2}$. For this we use equation \eqref{eq: min gamma}, and cancel the powers of $2$. Note that $z_k \le j$ since the received sequence can differ from a partial label by no more than the number $j$ of entries in the partial label. Since 
\begin{align}
    \sum^j_{z_k}\binom{j}{z_k}q^{z_k} p^{j-z_k}=1 \,,
\end{align}
then $\gamma \ge \frac{1}{2}(1+\Delta_{\max})$ for any $\Delta_{\max}$.
Using this, first we find the ``median'' posterior $\rho_{o_m}(y^{s_l})$, and compute a tentative $\Delta'_{\max}$ via \eqref{eq: lookahead constraint}. Since we need that $P_{\U_k} \ge 2^{-D_l}\gamma$, we also need $\delta_{k} \le 2^{-D_l}(1-\gamma)$ for each $k$, which leads to $\Delta'_{\max} \le (1-\gamma)$. Then we use $\gamma$ and  $\Delta'_{\max}$ to compute smallest $h$ that satisfies \eqref{eq: min gamma}, and then use $h$ to compute $\rho^{\min}_{o_\gamma}(y^{s_l+j})$ via equation \eqref{eq: min rho gamma k}. We repeat the process using the next posterior smaller than $\rho_{o_\gamma}(y^{s_l})$. If we obtain a larger $\rho^{\min}_{o_\gamma}(y^{s_l+j})$, we keep the larger value and repeat the process. If the next value of $\rho^{\min}_{o_\gamma}(y^{s_l+j})$ that we compute is smaller, we keep the previous one and stop the search.
Note that many posteriors share the same value, and since we keep then together in a group, as was done in \cite{9174232} and \cite{antonini2023}, we only test once each value of $\rho_{o_i}(y^{s_l}) \ge \rho_{o_m}(y^{s_l})$ that is shared by an entire group. This makes the search for an appropriate $\rho^{\min}_{o_\gamma}(y^{s_l+j})$  very fast. In practice, the complexity of this search is negligible compare to actually designing the partitions after finding $\rho^{\min}_{o_\gamma}(y^{s_l+j})$.

\section{Simulation Results}
\label{sec: simulation results}
\begin{figure}[t]
\centering
\ifCLASSOPTIONonecolumn
\includegraphics[width=0.8\textwidth]{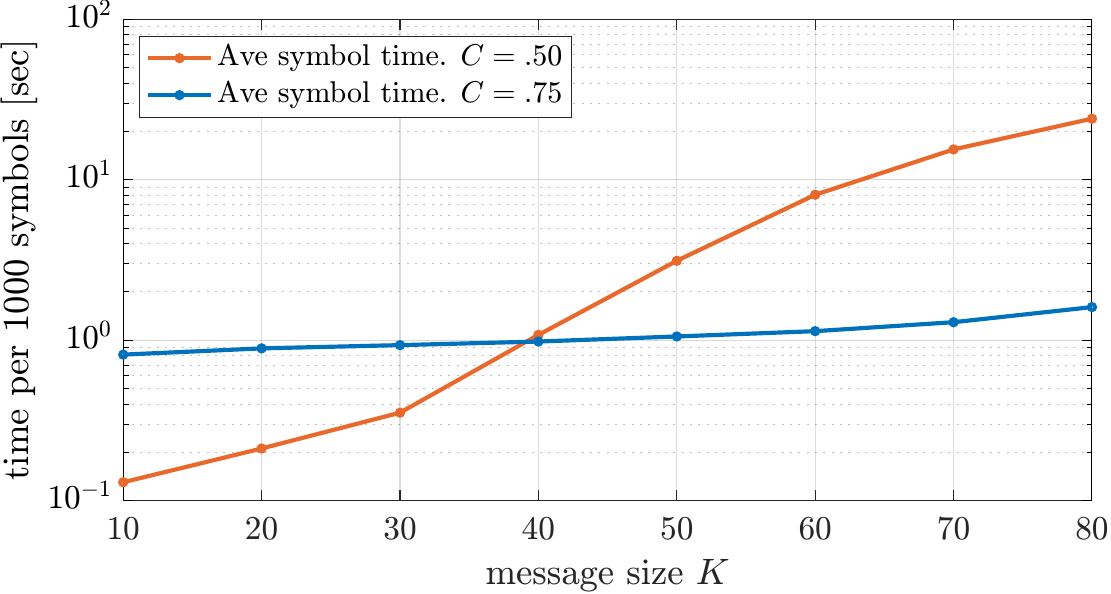}
\else
\includegraphics[width=0.5\textwidth]{Figures/LHtimeTxVsK.pdf}
\fi
\caption{Run-time complexity of the ``look-ahead'' algorithm vs. $K$, in average time per $1000$ symbols for channels with capacity $C=0.50$ and $C=0.75$. }
\vspace{-2.0mm}
\label{fig: symbol times}
\end{figure}
We implemented the ''look-ahead'' algorithm and obtained performance results to demonstrate how ``sparse'' the feedback times can be while maintaining a rate above the bounds for the non-sparse case. We show sparsity by the expected size $D_l$ of the ``blocks'' transmitted at each time $s_l\, l=1,2,\dots,\eta$.
The ``sparsity'' performance of the ``look-ahead'' algorithm is provided in Fig. \ref{fig: all sparse} as a function of message size $K$ for two channels with capacity $0.50$ and $0.75$. The solid line $-\circ$ shows the performance of the ``look-ahead'' in the communication phase, the target region where each $\rho_i(y^t) < 0.5$ where $2^{D_l}$ ``bins'' with $D_l > 1$ could be constructed and still satisfy constraint \eqref{eq: lookahead constraint}. For reference we show the overall $\E[D_l]$ including the systematic block $D_1=K$ and the average $\E[D_l\mid i \ge 2]$ that includes the times where $\exists i \in \Omega\, \rho_i(y^t) \ge 0.5$. Fig. \ref{fig: rate vs K} shows the rate performance of the ``look-ahead'' algorithm for the same simulations of Fig. \ref{fig: all sparse}, and the bounds \eqref{eq: stopping time optimized bound} and \eqref{eq: binomial tau} that validate the claim that the rate performance is above the bounds. The rate performance of the non-sparse algorithm in \cite{antonini2023} is provided for reference, which is no better than that of the ``look-ahead'' algorithm. 
The simulations show that as $K$ grows we can increase the sparsity, in the target region, up to an average $\E[D_l]$ of $5$ to $6$ bits per block.

The run-time complexity of the ``look-ahead'' algorithm as a function of channel crossover probability $p$ for $K=16,32, 64, 96$ is shown in Fig. \ref{fig: symbol times}. The complexity curves of the algorithm increases very rapidly with $p$ and with $K$. To the right the curves seem to taper down, but this is probably artifact introduced by a cap on the largest $D_l$, which we set at $D_l \le 12$ because of hardware memory restrictions.

\section{Conclusion}
\label{sec: conclusion}
This work explores how the frequency of feedback transmissions affects achievable rate when noiseless feedback of received symbols is used for posterior-matching communication.  Although previous works usually assume that each received symbol is fed back before the next transmission, this work shows that the frequency of the feedback can be significantly reduced with no noticeable loss in achievable rate.  No feedback is required until after the initial transmission of systematic bits.  After that,  careful partitioning allows multiple symbols to be transmitted before feedback is required for a new partitioning step.
\begin{figure}[t]
\centering
\ifCLASSOPTIONonecolumn
\includegraphics[width=0.8\textwidth]{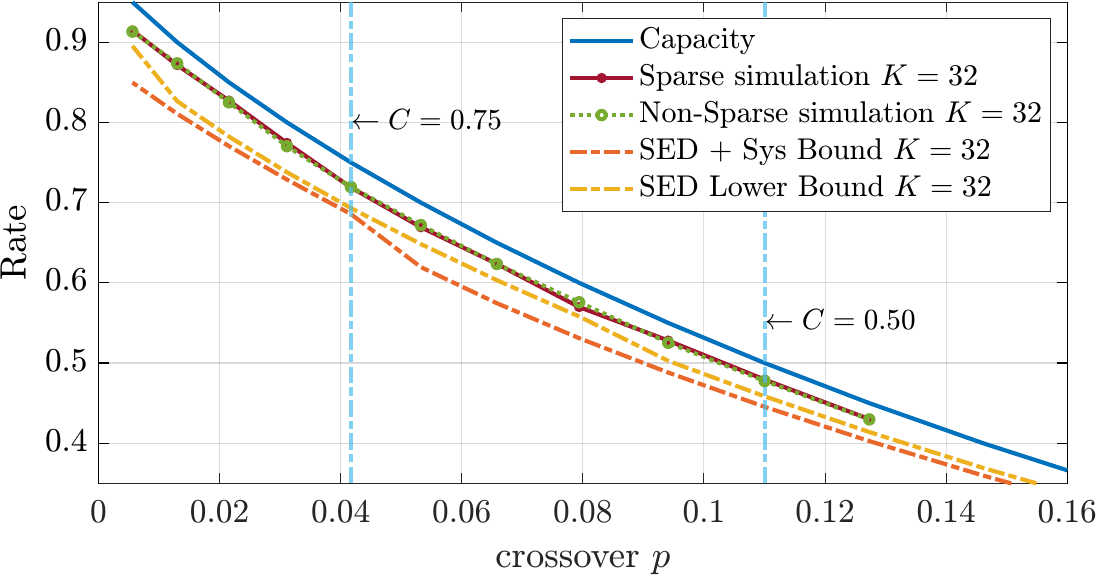}
\else
\includegraphics[width=0.5\textwidth]{Figures/LHrateVsCK32.pdf}
\fi
\caption{Rate performance vs. channel $p$  of the look-ahead algorithm for $K=32$. The solid solid dark blue curve shows the channel capacity. The ``look-ahead algorithm'' curve is the brown solid line $-\circ$. The green solid line $-\circ$ is for the non-sparse algorithm in \cite{antonini2023}. The orange line dash is the rate lower bound $K/\E[\tau]$ for systematic transmission using \eqref{eq: stopping time bound} and the yellow dash line is the lower bound from \eqref{eq: stopping time optimized bound} for uniform input distribution. }
\label{fig: rate vs C K32}
\vspace{-1.0em}
\end{figure}

\appendices
\label{appendix: appendices}
\section{Proof that constraint \eqref{eq: lookahead constraint} guarantees \eqref{eq: C step size} }
\label{sec: delta square proof}

Need to show that:
\begin{flalign}
    \Delta^2 \! \le \! \frac{2}{5}\rho_{o_m}(y^t) \! \implies \! E[U_\theta(t \! + \! 1) \! - \! U_\theta (t) | Y^t \! = \! y^t] \ge C \label{eq: delta proof req}&
\end{flalign}
To prove that \eqref{eq: delta proof req} holds, we will lower bound $E[U_\theta(t \! + \! 1) \! - \! U_\theta (t) | Y^t \! = \! y^t]$ by a function of $\Delta$ and $\rho_{o_m}(y^t)$ and then show that the lower bound satisfies \eqref{eq: delta proof req} whenever \eqref{eq: lookahead constraint} is satisfied.  
We will use intermediate expressions, where each lower bounds the previous expression and the last one only involves $\Delta$ and $\rho_{o_m}(y^t)$.
To start, we expand $E[U_\theta(t \! + \! 1) \! - \! U_\theta (t) | Y^t \! = \! y^t]$ using the definition.
\begin{align}
    E[&U_\theta (t+1)- U_\theta (t) | Y^t = y^t]
    \\
    &= \sum_{i\in \Omega}
    \rho_i(y^t)E[U_i(t \! + \! 1) \! - \! U_i(t) | Y^t \! = \! y^t, \theta \! = \! i] 
    \\
    &=\sum_{i\in S_0}
    \rho_i(y^t)E[U_i(t \! + \! 1) \! - \! U_i(t) | Y^t \! = \! y^t, \theta \! = \! i] \nonumber
    \\
    & \quad \quad + 
    \sum_{i\in S_1}
    \rho_i(y^t)E[U_i(t \! + \! 1) \! - \! U_i(t) | Y^t \! = \! y^t, \theta \! = \! i] 
\end{align}
The value $U_i(t)$ is a constant given $Y^t$. The next few steps consist of expressing the random variable $U_i(t+1)$ as $U_i(t)+C+f(Y^{t+1})$ to cancel $U_i(t)$ and $C$. Since $U_i(t+1)$ is defined in terms of $\rho_i(t+1)$, we first need $\rho_i(t+1)$.
From \cite{antonini2023} we have that:
\begin{flalign}
    \rho_i&(y^{t+1}) = \frac{\Pr(Y_{t+1}=y_t \mid \theta = i, Y^t = y^t)\rho_i(y^t)}{\sum_{j\in \Omega}\Pr(Y_{t+1}\mid j)\rho_j(y^t)}&
    \\
    &= \frac{\Pr(Y_{t+1}=y_t \mid \theta = i, Y^t = y^t)\rho_i(y^t)}
    {\underset{j\in S_{y_{t+1}}}{\sum\Pr(}Y_{t+1}\mid j)\rho_j(y^t)
    \underset{j\in \Omega \setminus S_{y_{t+1}}}{+\sum\Pr(Y_{t+1}} \! \mid j)\rho_j(y^t)} \, .&
\end{flalign}
Where $\Pr(Y_{t+1}\mid j)=\Pr(Y_{t+1}=y_{t+1}\mid Y^t=Y^t,  \theta = j)$. Since $X_{t+1} = \mathbbm{1}_{\theta \in S_1}$, the index of the set containing $\theta$, then 
\begin{alignat}{2}
    j\in &S_{y_{t+1}}& &\implies \Pr(Y_{t+1} \! = \! y_{t+1}\mid Y^t \! = \! Y^t,  \theta  \! = \! j)=q
    \nonumber
    \\
    j\in &\Omega \setminus S_{y_{t+1}}& 
    &\implies \Pr(Y_{t+1} \! = \! y_{t+1}\mid Y^t \! = \! Y^t,  \theta  \! = \! j)=p
    \nonumber
\end{alignat}
Then:
\begin{align}
    \rho_i&(y^{t+1})= \frac{\Pr(Y_{t+1}=y_t \mid \theta = i, Y^t = y^t)\rho_i(y^t)}{q \sum_{j\in S_{y_{t+1}}}\rho_j(y^t)+p\sum_{j\in \Omega \setminus S_{y_{t+1}}}\rho_j(y^t)}\label{eq: general update}
\end{align}
For $i\in S_0$ and $Y_{t+1} = 0$ we have:
\begin{flalign}
    \rho_i&(t \! + \! 1) = \frac{q\rho_i(y^t)}{P_0 q \! + \! P_1 p}
    =\frac{q\rho_i(y^t)}{\frac{1}{2} \! + \! \frac{\Delta(q-p)}{2}}
    =
    \frac{2q\rho_i(y^t)}{1 \! + \! \Delta(q \! - \! p)}
    \label{eq: S0 and Y0 update}\, .&
\end{flalign}
For $i\in S_0$ and $Y_{t+1} = 1$ we replace $q$ with $p$ in the top and the sign of $\Delta$ in the bottom. For $i\in S_1$ and $Y_{t+1} = 1$ we only change the sign of $\Delta$ in the bottom and for $i\in S_1$ and $Y_{t+1} = 0$ we only replace $q$ with $p$ in the top. 
Using the definition \eqref{eq: U process} of $U_i(t)$ we expand $\E[U_i(t\!+\!1) \mid Y^t \! = \! y^t, \theta \! = \! i]$,
for $i\in S_0$ to obtain:
\begin{flalign*}
    &\E[U_i(t\!+\!1) \! \mid \! Y^t, i]
    \! = \! q \log_2\frac{\frac{2q\rho_i(y^t)}{1 \! + \! \Delta(q \! - \! p)}}{1 \! - \! \frac{2q\rho_i(y^t)}{1 \! + \! \Delta(q \! - \! p)}} 
    \! + \! 
    p \log_2\frac{\frac{2p\rho_i(y^t)}{1 \! - \! \Delta(q \! - \! p)}}{1 \! - \! \frac{2p\rho_i(y^t)}{1 \! - \! \Delta(q \! - \! p)}}
    \, .&
\end{flalign*}

If $i\in S_1$ then, only the sign of $\Delta$ changes. 
Let $\iota_i$ be $1$ if $i \in S_0$ and $-1$ if $i \in S_1$, that is: $\iota_i = \mathbbm{1}_{i\in S_0}-\mathbbm{1}_{i\in S_1}$. Note that the numerators are $q\log_2(2q\rho_i(y^t))$ and $p\log_2(2p \rho_i(y^t)$, which we can reorganize as $1+q\log_2(q)+p\log_2(p)+\log_2(\rho_i(y^t)$, where the first three terms add up to $C$. When we subtract $U_i(t) = \log_2(\rho_i(y^t))-\log_2(1-\rho_i(y^t))$, a constant for fixed $Y^t$, the last term also vanishes and we only need to add $\log_2(1-\rho_i(y^t))$ to obtain:
\begin{flalign}
    \E[U_i(t \! &+ \! 1) \!  - \! U_i(t) \! \mid \! Y^t \! = \! y^t \! , \! \theta  \! = \! i] 
    \\
    &= \! q \log_2\frac{\frac{2q\rho_i(y^t)}{1 \! + \! \Delta(q \! - \! p)}}{1 \! - \! \frac{2q\rho_i(y^t)}{1 \! + \! \Delta(q \! - \! p)}} 
    \! + \! 
    p \log_2\frac{\frac{2p\rho_i(y^t)}{1 \! - \! \Delta(q \! - \! p)}}{1 \! - \! \frac{2p\rho_i(y^t)}{1 \! - \! \Delta(q \! - \! p)}}
    \\
    &= \! C \! + \! \log_2(\rho_i(y^t)) \! - \! U_i(t) \pm \log_2(1\!-\!\rho_i(y^t))\label{eq: add subtract rho}&
    \\
    &-q\log_2\left(1 \! - \! \rho_i(y^t) \! + \! (q \! - \! p)(\iota_i\Delta \! - \! \rho_i(y^t)\right)&
    \\
    &- p \log_2\left(1 \! - \! \rho_i(y^t) \! - \! (q \! - \! p)(\iota_i\Delta \! - \! \rho_i(y^t) \right)&
    \\
    &= C -q \! \log_2\left(1 \! + \! (q \! - \! p)\frac{ \iota_i\Delta \!  - \! \rho_i(y^t)}{1-\rho_i(y^t)}\right) \label{eq: 1-rho bottom q}&  
    \\
    &\quad \quad \quad - p \log_2\left(1 \! - \! (q \! - \! p)\frac{\iota_i\Delta \! - \! \rho_i(y^t)}{1 \! - \! \rho_i(y^t)}\right) \label{eq: 1-rho bottom p}&
    \\
    &\ge C - \log_2\left(1 + (q \! - \! p)^2\frac{ \iota_i\Delta
    -\rho_i(y^t)}{1-\rho_i(y^t)}\right) \, .& \label{eq: jensen over p q}
\end{flalign}
In \eqref{eq: add subtract rho} we add and subtract $\log_2(1\!-\!\rho_i(y^t))$ to recover  and cancel $U_i(t)=\log_2(rho_i(y^t))-\log_2(1-\rho_i(y^t))$. In \eqref{eq: 1-rho bottom q} and \eqref{eq: 1-rho bottom p}
we factor we use $q+p = 1$ divide the argument of the logs by $1\!-\!\rho_i(y^t)$ in $\log_2(1\!-\!\rho_i(y^t))$ and 
in \eqref{eq: jensen over p q} we use  Jensen's inequality over $p$ and $q$.
\ifCLASSOPTIONonecolumn
\begin{align}
    \sum_{i = 1}^M  \rho_i(y^t) \E[U_i(t+1)- & U_i(t)\mid Y^t, \theta = i] 
    \ge
    C - \sum_{i=1}^M \rho_i(y^t)\log_2\left(1 +(q-p)^2\frac{ \iota_i\Delta
    -\rho_i(y^t)}{1-\rho_i(y^t)}\right) \label{eq: sum rest}\, .
\end{align}
\else
\begin{flalign}
    \sum_{i = 1}^M  &\rho_i(y^t) \E[U_i(t+1)-U_i(t)\mid Y^t, \theta = i] \ge C\nonumber&
    \\
    &- \sum_{i=1}^M \rho_i(y^t)\log_2\left(1 +(q-p)^2\frac{ \iota_i\Delta
    -\rho_i(y^t)}{1-\rho_i(y^t)}\right) \, .&
    \label{eq: sum rest}
\end{flalign}
\fi
The rest of the proof consist showing that the sum
in \eqref{eq: sum rest} is non-negative when $\Delta$ satisfies rule \eqref{eq: lookahead constraint}. We start by lower bounding the sum with an expression independent of $i$. 
Let $\delta$ be twice the amount that the $c.d.f.$, evaluated at the median $\rho_{o_m}(y^t)$, exceeds half, and let $R \in [0,1]$ be the fraction of posteriors greater than $\rho_{o_m}(y^t)$ that are assigned to $S_0$:
\begin{align}
    \frac{1}{2} &\le \sum_{i=1}^{m}\rho_{o_i}(y^t) = \frac{1+\delta}{2} \le  \frac{1+\rho_{m}(y^t)}{2}
    \\
    R\frac{1+\delta}{2} &=\sum_{i \in S_0}\rho_{i}(y^t)\mathbbm{1}_{[\rho_i(y^t)\ge\rho_{o_m}(y^t)]} 
    \,.
\end{align}
Then the sum of posteriors at or above $\rho_{o_m}(y^t)$ is at least $\frac{1+\delta}{2}$. 
Now we proceed to lower bound the sum in \eqref{eq: sum rest} by expressions that only depend on $\Delta, \rho_{o_m}(y^t), \delta, R$. There are two expressions to consider, first $\frac{ \Delta-\rho_i(y^t)}{1-\rho_i(y^t)}$ for the case that $\iota = 1$ and second $\frac{ -\Delta-\rho_i(y^t)}{1-\rho_i(y^t)}$ for the case that $\iota = -1$. Depending on whether $i\in S_0$ or $i\in S_1$ and whether $\rho_i(y^t)$ is at least $\rho_{o_m}(y^t)$ or smaller than $\rho_{o_m}(y^t)$, one of the following four bounds will apply:
\begin{alignat}{5}
    &\rho_i(y^t)& &\ge \rho_{o_m}(y^t)& &\implies& \frac{\Delta-\rho_{o_m}(y^t)}{1-\rho_{o_m}(y^t)} &\ge
    \frac{\Delta-\rho_i(y^t)}{1-\rho_i(y^t)} \label{eq: S0 and SR}&
    \\
    &\rho_i(y^t)& &\ge \rho_{o_m}(y^t)& &\implies& \frac{- \Delta \! - \! \rho_{o_m}(y^t)}{1-\rho_{o_m}(y^t)} &\ge
    \frac{-\Delta-\rho_i(y^t)}{1-\rho_i(y^t)}\label{eq: S1 and SR}&
    \\
    &\rho_i(y^t)& &< \rho_{o_m}(y^t)& &\implies& \Delta &\ge \frac{\Delta-\rho_i(y^t)}{1-\rho_i(y^t)} \label{eq: S0 and not SR}&
    &
    \\
    &\rho_i(y^t)& &< \rho_{o_m}(y^t)& &\implies& -\Delta &\ge  \frac{-\Delta-\rho_i(y^t)}{1-\rho_i(y^t)}\label{eq: S1 and not SR}&
\end{alignat}
We will use the following posteriors sum for the four cases:
\begin{flalign}
    \Pr\{i \! \in \! S_0 \!: \rho_i(y^t)\ge \rho_{o_m}(y^t)\} =&R\frac{1+\delta}{2} \label{eq: S0 and R}&
    \\
    \Pr\{i \! \in \! S_0 \!: \rho_i(y^t) < \rho_{o_m}(y^t)\} 
    = &\frac{1+\Delta}{2}-R\frac{1+\delta}{2} \label{eq: S0 and not R}&
    \\
    \Pr\{i \! \in \! S_1 \!: \rho_i(y^t)\ge \rho_{o_m}(y^t)\}
    = &(1-R)\frac{1+\delta}{2} \label{eq: S1 and 1-R}&
    \\
    \Pr\{i \! \in \! S_1 \!: \rho_i(y^t) < \rho_{o_m}(y^t)\}
    = &\frac{1\!-\!\Delta}{2}\!-\!
    (1\!-\!R)\frac{1\!+\!\delta}{2} \label{eq: S1 and not 1-R}&
\end{flalign}
For the set in \eqref{eq: S0 and R} we use bound \eqref{eq: S0 and SR}, for the set in \eqref{eq: S0 and not R} we use bound \eqref{eq: S1 and SR}, for the set in \eqref{eq: S1 and 1-R} we use bound \eqref{eq: S0 and not SR}, and for the set in \eqref{eq: S1 and not 1-R} we use bound \eqref{eq: S1 and not SR}. The sum in \eqref{eq: sum rest} is lower bounded by: 
\begin{flalign}
    -\!\sum_{i=1}^M &\rho_i(y^t)\log_2\left(1 +(q-p)^2\frac{ \iota_i\Delta
    -\rho_i(y^t)}{1-\rho_i(y^t)}\right) &
    \\
    \ge
    &-R\frac{1+\delta}{2}\log_2\left(1 +(q-p)^2\frac{\Delta-\rho_{o_m}(y^t)}{1-\rho_{o_m}(y^t)}\right)&
    \\
    &-\left(\frac{1+\Delta}{2}-R\frac{1+\delta}{2}\right)
    \log_2\left(1 +(q-p)^2\Delta \right)&
    \\
    &-\!(1\!-\!R)\frac{1\!+\!\delta}{2}\log_2\!\left(\!1\! + \!(q\!-\!p)^2\frac{- \Delta \! - \! \rho_{o_m}(y^t)}{1\!-\!\rho_{o_m}(y^t)}\!\right)&
    \\
    &-\left(\frac{1\!-\!\Delta}{2}\!-\!
    (1\!-\!R)\frac{1\!+\!\delta}{2}\right)\log_2\left(1\! -\!(q\!-\!p)^2\Delta\right)&
    \\
    &\ge -\log_2\left(1 - \frac{(q-p)^2}{2}f(\Delta, R, \delta, \rho_{o_m}(y^t))\right) \, .&
    \label{eq: Jensen on Delta R and delta}
\end{flalign}
Where the last inequality \eqref{eq: Jensen on Delta R and delta} follows by applying Jensen's inequality over the ``weights'' in \cref{eq: S0 and R,eq: S0 and not R,eq: S1 and 1-R,eq: S1 and not 1-R}. Collecting the terms with $\Delta$ and with $\frac{\pm \Delta - \rho_{o_m}(y^t)}{1-\rho_{o_m}(y^t)}$ we obtain:
\begin{align}
    &\Delta (-(1\!+\!\Delta)\!+\!R(1\!+\!\delta)\!+\!(1\!-\!\Delta)\!-\!(1\!-\!R)(1\!+\!\delta))
    \\
    &=
    \Delta ((1-1-\Delta-\Delta)-(1-R-R)(1+\delta))
    \\
    &=
    \Delta (-2\Delta+(1-2R)(1+\delta))
    \\
    &\frac{1\!+\!\delta}{1-\rho_{o_m}}(R(\rho_{o_m}(y^t)\!-\!\Delta)\!+\!(1\!-\!R)(\!\Delta\!+\!\rho_{o_m}(y^t))) 
    \\
    &=
    \frac{\Delta(1+\delta)(1-2R)}{1-\rho_{o_m}(y^t)}
    +
    \frac{\rho_{o_m}(y^t)(1+\delta)(1-R+R)}{1-\rho_{o_m}(y^t)}
    \\
    &=
    \frac{(1+\delta)(\Delta(1-2R)+\rho_{o_m}(y^t))}{1-\rho_{o_m}(y^t)}\\
    f(\Delta, &R, \delta, \rho_{o_m}(y^t))= \frac{\Delta(1\!-\!2R)(1\!+\!\delta)+\rho_{o_m}(y^t)(1\!+\!\delta)}{1\!-\!\rho_{o_m}(y^t)} \nonumber&
    \\
    &\quad -\Delta (2\Delta-(1-2R)(1+\delta)) 
    \, .\label{eq: f from Jensen}
\end{align}

To show that rule \eqref{eq: lookahead constraint} suffices to satisfy inequality \eqref{eq: C step size} we only need to show that it guarantees that $f(\Delta, R, \delta, \rho_{o_m}(y^t))$ is non negative.
We only need to show that:
\begin{align*}
    \rho_{o_m}(y^t)(1\!+\!\delta)
    &\ge 2\Delta^2(1\!-\!\rho_{o_m}(y^t))\!- \!\Delta\rho_{o_m}(y^t)(1\!-\!2R)(1\!+\!\delta)
\end{align*}
We want to construct a bound that does not depend on which items are in $S_0$ and $S_1$, thus we consider the worst case scenario value of $\Delta(1-2R)$ and remove the dependence on $R$. For $\Delta > 0$ this happens at $R=1$ and for $\Delta \le 0$ at $R=0$. Then the expression with $\Delta$ is always negative. Let $\alpha = |\Delta|$ and set $\Delta(1-2R)=\alpha$. We need:
\begin{align}
    &\rho_{o_m}(y^t)(1+\delta)(1-\alpha)
    \ge 2\alpha^2(1-\rho_{o_m}(y^t)) 
\end{align}
Since $0 \le \delta \le 1$, and $0 < (1-\alpha) < 1$, let $\delta=0$ for a bound:
\begin{align}
    \rho_{o_m}(y^t)(1-\alpha)
    &\ge 2\alpha^2(1-\rho_{o_m}(y^t)) 
\end{align}
Let $\alpha^2$ be bounded by a linear function of $\rho_{o_m}(y^t)$ of the form $\alpha^2 \le \frac{a}{2b}\rho_{o_m}(y^t)$, for some $0 \le \frac{a}{2b} < 1$, we also need:
\begin{align}
    \rho_{o_m}(y^t)(1-\alpha)
    &\ge \frac{a}{b}\rho_{o_m}(y^t)(1-\rho_{o_m}(y^t)) 
    \label{eq: bound alpha square}
    \\
    \alpha &\le \frac{b-a}{b}+\frac{a}{b}\rho_{o_m}(y^t)
    \label{eq: bound alpha}
\end{align}
To complete the proof we need to show that inequalities \cref{eq: bound alpha square,eq: bound alpha} hold for any $\alpha^2 \le \frac{a}{2b}\rho_{o_m}(y^t)$ with $a=4$ and $b=5$,  where $\frac{a}{2b} = \frac{2}{5}$. For these $a,b$ we have:
\begin{align}
    \alpha &\le \frac{1}{5}+\frac{4}{5}\rho_{o_m}(y^t)
\end{align}
Let $\rho_{o_m}(y^t) \le \frac{1}{10}$ then  $\alpha^2 \le \frac{2}{5}\rho_{o_m}(y^t)\implies \alpha \le \frac{1}{5}$.
Now let $\frac{1}{10} \le \rho_{o_m}(y^t) \le \frac{49}{250}$ so that $\alpha^2 \le \frac{49}{625} \implies \alpha \le \frac{7}{25}$. On the right side pick the smallest $\rho_{o_m}(y^t)=\frac{1}{10}$ to obtain:
\begin{align}
    \alpha &\le \frac{1+4\frac{1}{10}}{5}=\frac{7}{25}
\end{align}
We can show \cref{eq: bound alpha square,eq: bound alpha} holds for any $\rho_{o_m}(y^t)\in [0,1]$ by repeating the previous step with  $\rho_{o_m}(y^t)\in [r_j,r_{j+1}]$ for $r_j=\frac{49}{250},\frac{5}{18},\frac{2}{5}, \frac{5}{8}, \frac{3}{4}, 1$.
This completes the proof that $\alpha^2 \le \frac{2}{5}\rho_{o_m}(y^t)$ suffices to satisfy inequality \eqref{eq: C step size}. 

\section{Proof of inequality \eqref{eq: update bound}}
\label{sec: rho tsi min}
To proof \eqref{eq: update bound} we start we transform the expression for $\rho_i(y^{s_l\!\! + \! j})$ into a function of $\rho_i(y^{s_l})$, the Hamming distance $z_{s_l\!,j}(i)$ between the label of the bin containing $i$ and the feedback symbol $y^{s_{l}\!+\!j}_{s_{l}\!+\!1}$ and the bin posteriors at time $t=s_l$ and the bin probabilities $\rho_{\U_k}(y^{s_l})$. Then we lower bound the expression with an upper bound on the bin probabilities in the denominator as follows:
\begin{flalign}
    \rho_i(y^{s_l\!\! + \! j})  
    &= 
    \Pr( \theta \! = \! i \! \mid \! Y^{s_{l}\!+\!j}_{1} \! \! = \! y^{s_{l}\!+\!j}_{1} )
    = \frac{\Pr( \theta \! = \! i, Y^{s_{l}\!+\!j}_{1} \! \! = \! y^{s_{l}\!+\!j}_{1} ) }{\Pr(Y^{s_{l}\!+\!j}_{1} \! \! = \! y^{s_{l}\!+\!j}_{1} ) }
    \nonumber &
    \\
    &= \! \frac{\Pr(Y^{s_{l}\!+\!j}_{s_{l}+1} \! = \! y^{s_{l}\!+\!j}_{s_{l}\!+\!1}  \mid \! Y^{s_l} \! = \! y^{s_l}, \! \theta \! = \! i)\rho_i(y^{s_l})}{\underset{r\in \Omega}{\sum}\!\! \Pr(Y^{s_{l}\!\!+\!j}_{s_{l}\!+\! 1} \! = \! y^{s_{l}\!+\!j}_{s_{l}\!+\!1} \! \mid \! Y^{s_l} \! = \! y^{s_l} \! , \! \theta \! = \! r)\rho_r(y^{s_l}\!)} \nonumber &
    \\ 
    &=\frac{q^{j-z_{s_l\!,j}(i)}
    p^{z_{s_l,j}(i)} \rho_{i}(y^{s_l})}
    {\overset{2^{D_l} \!- \! 1}{\underset{k=0}{\sum}}
    q^{j-z_{s_l\!,j}(\U_k)}
    p^{z_{s_l,j}(\U_k)} \rho_{\U_k}(y^{s_l})} \label{eq: bin update}&
    \\
    &=\frac{q^{j-z_{s_l\!,j}(i)}
    p^{z_{s_l,j}(i)} \rho_{i}(y^{s_l})}
    {\overset{2^{D_l} \!- \! 1}{\underset{k=0}{\sum}}
    q^{j-z_{s_l\!,j}(\U_k)}
    p^{z_{s_l,j}(\U_k)} (2^{-D_l}+\delta_k)}
    \label{eq: replace delta k}&
    \\
    &\ge 
    \frac{q^{j-z_{s_l\!,j}(i)}
    p^{z_{s_l,j}(i)} \rho_{i}(y^{s_l})}
    {\overset{2^{D_l} \!- \! 1}{\underset{k=0}{\sum}}
    q^{j-z_{s_l\!,j}(\U_k)}
    p^{z_{s_l,j}(\U_k)} 2^{-D_l}(1 \! + \! \Delta_{\max
    })}
    \label{eq: bound delta k} &
    \\
    &=
    \frac{q^{j-z_{s_l\!,j}(i)}
    p^{z_{s_l,j}(i)} \rho_{i}(y^{s_l})}
    {\overset{D_l}{\underset{k=0}{\sum}}\binom{j}{k}
    q^{j-k}
    p^{k} 2^{-j}(1 \! + \! \Delta_{\max})} 
    &
    \\
    &=
    \frac{2^{j} q^{j-z_{s_l\!,j}(i)}
    p^{z_{s_l,j}(i)} \rho_{i}(y^{s_l})}
    {1 \! + \! \Delta_{\max}} 
    \, . 
\end{flalign}
In \eqref{eq: replace delta k} we have used the definition of $\delta_k=\rho_{\U_k}(y^{s_l})-2^{-D_l}$ and \eqref{eq: bound delta k} follows from \eqref{eq: max delta}: via $2^{-D_l}\Delta_{\max} = \max_j\{\delta_j\}\le \delta_k$.

\ifCLASSOPTIONcaptionsoff
  \newpage
\fi

\bibliographystyle{IEEEtran}
\bibliography{IEEEabrv,references}
\end{document}